\documentclass[12pt,preprint]{aastex}

\usepackage{epsfig}

\newcommand{\begeq}{\begin{equation}}
\newcommand{\fineq}{\end{equation}}

\newcommand{\rs}{r_{_{\rm S}}}
\newcommand{\pseudophi}{\Phi}
\newcommand{\OmegaK}{\Omega_{\rm K}}
\newcommand{\ellK}{\ell_{\rm K}}

\newcommand{\vff}{v_{\rm ff}}
\newcommand{\Qjet}{L_{\rm jet}}

\newcommand{\SgrA}{Sgr\,A$^*$}
\newcommand{\gapprox}{\lower.4ex\hbox{$\;\buildrel
>\over{\scriptstyle\sim}\;$}}
\newcommand{\lapprox}{\lower.4ex\hbox{$\;\buildrel
<\over{\scriptstyle\sim}\;$}}

\slugcomment{accepted by ApJ}

\shorttitle{Shocks in Viscous, Advection-Dominated Disks}

\shortauthors{Das, Becker, \& Le}

\begin{document}

\title{DYNAMICAL STRUCTURE OF VISCOUS ACCRETION DISKS WITH SHOCKS}

\author{Santabrata Das}

\affil{Korea Astronomy and Space Science Institute,
61-1, Hwaam Dong, Yuseong-Gu, Daejeon 305 348, South Korea; sbdas@kasi.re.kr}

\author{Peter A. Becker\altaffilmark{1}}

\affil{Department of Computational and Data Sciences, George Mason
University, Fairfax, VA 22030-4444, USA; pbecker@gmu.edu}

\and

\author{Truong Le}

\affil{Space Telescope Science Institute, 3700 San Martin Drive,
Baltimore, MD 21218, USA; tle@stsci.edu}

\altaffiltext{1}{Also at: Department of Physics and Astronomy, George
Mason University, Fairfax, VA 22030-4444, USA.}

\begin{abstract}
We develop and discuss global accretion solutions for viscous ADAF disks
containing centrifugally supported isothermal shock waves. The fact that
such shocks can exist at all in ADAF disks is a new result.
Interestingly, we find that isothermal shocks can form even when the
level of viscous dissipation is relatively high. In order to better
understand this phenomenon, we explore all possible combinations of the
fundamental flow parameters, such as specific energy, specific angular
momentum, and viscosity, to obtain the complete family of global
solutions. This procedure allows us to identify the region of the
parameter space where isothermal shocks can exist in viscous ADAF disks.
The allowed region is maximized in the inviscid case, and it shrinks as
the level of viscous dissipation increases. Adopting the canonical value
$\gamma=1.5$ for the ratio of specific heats, we find that the shock
region disappears completely when the Shakura-Sunyaev viscosity
parameter $\alpha$ exceeds the critical value $\sim 0.27$. This
establishes for the first time that steady ADAF disks containing shocks
can exist even for relatively high levels of viscous dissipation. If an
isothermal shock is present in the disk, it would have important
implications for the acceleration of energetic particles that can escape
to power the relativistic jets commonly observed around underfed,
radio-loud black holes. In two specific applications, we confirm that
the kinetic luminosity lost from the disk at the isothermal shock
location is sufficient to power the observed relativistic outflows in
M87 and \SgrA.
\end{abstract}

% The different journals have different requirements for keywords.  The
% keywords.apj file, found on aas.org in the pubs/aastex-misc directory,
% contains a list of keywords used with the ApJ and Letters.  These are
% usually assigned by the editor, but authors may include them in their
% manuscripts if they wish.

\keywords{accretion, accretion disks --- hydrodynamics --- black hole physics
--- galaxies: jets.}

\section{INTRODUCTION}

The accretion of matter onto compact objects is the fundamental
mechanism powering a variety of high-energy astrophysical sources, such
as low-mass X-ray binaries, massive black holes, and active galactic
nuclei. Shakura \& Sunyaev (1973) and Novikov \& Thorne (1973) laid the
foundation for our understanding of these sources by developing the
first physical models for geometrically thin accretion disks. These
early models did not treat the pressure and advection terms in the
conservation equations correctly, and furthermore no attempt was made to
satisfy the inner boundary conditions at the event horizon. Instead, the
disk was simply terminated at the marginally stable orbit. The inner
regions of these disks were shown to be viscously and thermally unstable
by Lightman \& Eardley (1974), and the model was subsequently improved
by Paczy\'nski \& Bisnovatyi-Kogan (1981) and Muchotrzeb \& Paczy\'nski
(1982) who incorporated advection and pressure effects into their
models. Following a similar approach, the general global solution for
accretion in the hydrodynamic limit, including advection, viscosity, and
thermal effects, was obtained by Chakrabarti (1990, 1996).

\subsection{Advection-Dominated Accretion}

The ADAF model remains a central paradigm in contemporary black-hole
accretion theory as a possible explanation for the dynamical structure
of sources with significantly sub-Eddington accretion rates (see Yuan
2007 for a review). The gas in the inner regions of such disks is quite
tenuous, and therefore the plasma radiates very inefficiently, leading
to high temperatures. At large radii, the hot ADAF inner region is
expected to merge with a cool, geometrically thin Shakura \& Sunyaev
outer region characterized by a Keplerian angular momentum profile.
Since any global accretion solution onto a black hole must cross the
event horizon supersonically in order to satisfy the fundamental
boundary conditions imposed by general relativity, ADAF accretion disks
are necessarily transonic. Narayan, Kato, \& Honma (1997) studied the
behavior of transonic ADAF disks subject to the Shakura-Sunyaev
viscosity prescription. The associated asymptotic boundary conditions
governing the dynamics close to the event horizon, and the related
physical restrictions on the viscosity prescription, were examined by
Becker \& Subramanian (2005) and Becker \& Le (2003). Interest in the
ADAF model remains strong, and it has been the focus of many recent
studies. For example, Narayan \& McClintock (2008) have
analyzed the properties of ADAF disks close to the event horizon;
Takahashi (2007) has explored the influence of causal viscosity
prescriptions on the structure of ADAF disks; Ma, Yuan, \& Wang (2007)
have probed the role of magnetically-induced torques in ADAF disks; and
Yuan, Ma, \& Narayan (2008) have examined the properties of a simplified
global model that utilizes an approximate form for the radial momentum
equation.

\subsection{Critical Structure of Transonic Disks}

Due to the supersonic nature of the inflow at the event horizon, any
physically achievable flow configuration must contain at least one
critical point, where the flow transitions from subsonic to supersonic.
However, for a given set of incident flow parameters (e.g., specific
energy, specific angular momentum and viscosity), the flow may possess
multiple critical points (Chakrabarti 1989a, 1989b, 1990, 1996;
Chakrabarti \& Das 2004; Le \& Becker 2005). This suggests the possible
existence of global flow solutions with a standing shock wave located
between two critical points. In their early study of viscous, transonic
ADAF disks, Narayan et al. (1997) focused on global solutions that pass
through a single critical point, and consequently they did not obtain
any shocked disk solutions. The problem of the structure of ADAF disks
governed by the Shakura-Sunyaev viscosity prescription was recently
reexamined by Becker, Das, \& Le (2008), who demonstrated that
isothermal shocks can exist in such disks even in the presence of
substantial viscosity.

In the present paper, we extend the approach taken by Becker, Das, \&
Le (2008) by performing a complete analysis of the structure of viscous
ADAF disks containing either single or multiple critical points.
Solutions with multiple critical points may include a standing shock, if
it is possible to satisfy the shock jump conditions at some radius $r_*$
between the outer critical point (at radius $r=r_c^{\rm out}$) and the
inner critical point (at radius $r=r_c^{\rm in}$). Shocks occur when the
accretion flow is impeded due to the presence of a ``centrifugal
barrier'' situated close to the horizon. Although such barriers can
exist even in viscous disks, they begin to dissipate as the level of
viscosity increases because of the increasing efficiency of the outward
diffusion of angular momentum. Hence disks with large viscosity
parameters are not expected to possess shocks, and our results confirm
this. In the post-shock region, the flow accelerates towards the black
hole, eventually passing through the inner critical point and crossing
the horizon supersonically.

\subsection{Shock Waves in Viscous Disks}

The question of the stability of viscous accretion flows containing
standing shock waves is currently not resolved. However, an important
new development in the field was provided by Nagakura \& Yamada (2008),
who used both linear stability analysis and general relativistic
hydrodynamical simulations to investigate the stability properties of
shock waves in inviscid, advection-dominated accretion flows around
Schwarzschild black holes. When subjected to strong perturbations around
the initial condition, the authors find that the dynamical configuration
is stable, in the sense that the shock continues to exist, although the
shock radius may oscillate if the perturbations are non-axisymmetric.
Similar results were obtained by Okuda et al. (2007). Although these
authors do not consider the effect of viscosity in their work, their
conclusions nonetheless support the general idea that shocks may be able
to exist in a stable configuration in relativistic disks. Consequently
their findings provide a sound theoretical basis for the study of the
structure of shocked, viscous disks developed in the present paper. When
a shocked solution is dynamically possible, we argue based on the second
law of thermodynamics that a shock will form since shocked-disk
solutions possess higher entropy than smooth solutions (Becker \&
Kazanas 2001).

The study of shock waves in accretion flows around compact objects has
been undertaken by a variety of authors using either analytical methods
(Fukue 1987; Chakrabarti 1989a, 1989b, 1990, 1996; Abramowicz \&
Chakrabarti 1990; Nobuta \& Hanawa 1994; Yang \& Kafatos 1995; Yuan,
Dong, \& Lu 1996; Caditz \& Tsuruta 1998; Kovalenko \& Lukin 1999; Das,
Chattopadhyay, \& Chakrabarti 2001; Das \& Chakrabarti 2004; Le \&
Becker 2004, 2005, 2007; Yu, Lou, Bian, \& Wu 2006; Becker, Das, \& Le
2008) or numerical simulation (Hawley et al. 1984a, 1984b; Chakrabarti
\& Molteni 1993; Molteni, Sponholz, \& Chakrabarti 1996; Ryu,
Chakrabarti \& Molteni 1997; Chakrabarti, Acharyya, \& Molteni 2004).
The physical configuration of the shock depends on the detailed
microphysics of the dissipation mechanism operating at the shock. The
three types of shocks most frequently studied are (1) Rankine-Hugoniot
shocks, (2) isentropic shocks, and (3) isothermal shocks (Chakrabarti
1990; Abramowicz \& Chakrabarti 1990). Rankine-Hugoniot shocks possess
the largest post-shock temperatures because the energy flux is strictly
conserved as the gas crosses the shock. Hence shocks of this kind are
radiatively very inefficient. In an isentropic shock, the entropy
generated via dissipation is immediately radiated away, and consequently
such shocks display smaller temperature increases than the
Rankine-Hugoniot variety. In an isothermal shock, the temperature does
not increase at all because all of the dissipated energy and entropy
escapes from the flow at the shock location. Hence isothermal shocks
represent the most efficient mechanism for removing energy from the
flow.

Le \& Becker (2004, 2005, 2007) demonstrated that the energy removed
from the flow at the isothermal shock location can be understood
physically as a consequence of the first-order Fermi acceleration of
relativistic particles in the disk, combined with the diffusion and
escape of the accelerated particles through the upper and lower surfaces
of the disk. This approach facilitates the development of a completely
self-consistent, coupled model that simultaneously describes the
dynamical structure of the disk/shock system and also the energy
spectrum of the relativistic particles escaping from the disk to form
the observed jet outflows. We therefore focus on isothermal shocks in
the present paper.

The properties of isothermal shock waves in accretion disks have been
studied previously by Abramowicz \& Chakrabarti (1990), Lu \& Yuan
(1998), Fukumura \& Tsuruta (2004), and Das, Pendharkar, \& Mitra
(2003). However, the studies cited above all focused on inviscid flow
models, which probably do not adequately describe real accretion disks.
In the present paper, we obtain for the first time detailed dynamical
solutions for the structure of ADAF disks containing isothermal shocks.
The remainder of the paper is organized as follows. In \S~2 we discuss
the ADAF model assumptions, the governing equations, and the critical
conditions for the structure of ADAF disks. We discuss the isothermal
shock jump conditions and examine global accretion solutions in \S~3,
and we explore the properties of the standing shock and classify the
parameter space for shocked accretion solutions in \S~4. Our dynamical
formalism is used to model the development of the relativistic outflows
in M87 and \SgrA in \S~5, and concluding remarks are presented in \S~6.

\section{GOVERNING EQUATIONS}

We consider a thin, axisymmetric, viscous advection-dominated accretion
disk structure. The disk is tenuous because it accretes matter at well
below the Eddington rate, and therefore energy losses due to radiative
processes are inefficient. General relativistic effects are taken into
account in an approximate manner by utilizing the pseudo-Newtonian
potential introduced by Paczy\'nski \& Wiita (1980). The self-gravity of
the disk is neglected. Under these assumptions, there are two conserved
quantities in viscous ADAF disks, namely the mass transport rate
\begeq
\dot M = 4 \pi r H \rho \, v \ ,
\label{eq1}
\fineq
and the angular momentum transport rate
\begeq
\dot J = \dot M r^2 \Omega - \mathcal{G} \ ,
\label{eq2}
\fineq
where $r$ is the radial coordinate, $H$ denotes the disk half-thickness,
$\rho$ is the mass density, $v$ is the (positive) inflow velocity,
$\Omega$ is the angular velocity, and ${\cal G}$ denotes the torque. The
energy transport rate is given by
\begeq
\dot E = - \mathcal{G} \, \Omega +
\dot M\left({1 \over 2} \, v^2 + {1 \over 2} \, v^2_{\phi}
          + {P+U \over \rho} + \pseudophi \right) \ ,
\label{eq3}
\fineq
where $v_{\phi}=r \, \Omega$ denotes the azimuthal velocity, $P =
(\gamma-1)\,U$ is the pressure, $U$ is the internal energy density, and
$\Phi$ represents the pseudo-Newtonian potential, defined by
(Paczy\'nski \& Wiita 1980)
\begeq
\Phi(r) \equiv - {GM \over r-\rs} \ ,
\label{eq4}
\fineq
for a black hole with mass $M$ and Schwarzschild radius $\rs = 2GM/c^2$.
This potential successfully reproduces many aspects of the spacetime
geometry around a nonrotating black hole. We assume that the adiabatic
index, $\gamma$, maintains a constant value throughout the flow, and the
transport rates $\dot M$, $\dot J$, and $\dot E$ are all defined to be
positive for inflow. Under our assumptions, $\dot M$ and $\dot J$ are
conserved throughout the disk, and therefore they represent the rates at
which mass and angular momentum enter the black hole, respectively. In
general, the energy transport rate $\dot E$ is also conserved, except at
the location of an isothermal shock if one is present in the flow.

The torque ${\cal G}$ is associated with the gradient of the angular
velocity $\Omega$ through the standard relation (e.g., Frank, King, \&
Raine 2002)
\begeq
\mathcal{G}(r) =
- 4 \pi r^3 H \rho \, \nu {d\Omega \over dr} \ ,
\label{eq5}
\fineq
where $\nu$ is the kinematic viscosity. In the present work, we adopt
the Shakura-Sunyaev (1973) viscosity prescription, 
\begeq
\nu(r) = {\alpha a^2 \over \OmegaK} \ ,
\label{eq6}
\fineq
where $\alpha$ is a constant, $\OmegaK$ denotes the Keplerian angular
velocity, defined by
\begeq
\OmegaK^2(r) \equiv {GM \over r(r-\rs)^2}
= {1 \over r} {d\pseudophi \over dr} \ ,
\label{eq7}
\fineq
and $a$ denotes the isothermal sound speed, defined by
\begeq
a^2(r) = {P \over \rho} \ .
\label{eq8}
\fineq

Combining equations~(\ref{eq1}), (\ref{eq2}), (\ref{eq5}), and
(\ref{eq6}), we find that the gradient of the angular velocity can be
written as
\begeq
{d\Omega \over dr} = -{v \OmegaK (\ell - \ell_0) \over
\alpha r^2 a^2} \ ,
\label{eq9}
\fineq
where $\ell \equiv r^2 \Omega$ denotes the angular momentum per unit
mass for the accreting gas at radius $r$, and $\ell_0 \equiv \dot J/\dot
M$ is the angular momentum transport rate per unit mass. Since the
torque $\cal G$ vanishes at the event horizon (Becker \& Le 2003), it
follows from equation~(\ref{eq2}) that $\ell_0$ is also equal to the
specific angular momentum of the material entering the black hole. The
radial derivative of $\ell$ is given by
\begeq
{d\ell \over dr} = 2 \, r \, \Omega + r^2 {d\Omega\over dr} \ ,
\label{eq10}
\fineq  
which can be combined with equation~(\ref{eq9}) to obtain the
differential equation
\begeq
{d\ell \over dr} = {2 \, \ell \over r} - {v \ellK (\ell - \ell_0)
\over \alpha a^2 r^2} \ ,
\label{eq11}
\fineq
where the Keplerian specific angular momentum $\ellK$ is defined by
\begeq
\ellK(r) \equiv r^2 \OmegaK(r) = {r^{3/2} \sqrt{GM} \over r-\rs}
\ .
\label{eq12}
\fineq

The disk half-thickness $H$ is estimated using the standard vertical
hydrostatic relation
\begeq
H(r) = {a r^2\over \ellK} \ .
\label{eq13}
\fineq
In a steady state, the comoving radial acceleration rate in the frame
of the accreting gas can be written as
\begeq
{Dv \over Dt} \equiv -v {dv \over dr} = {1 \over \rho}
{dP \over dr} + {\ellK^2 - \ell^2 \over r^3} \ .
\label{eq14}
\fineq
By eliminating the torque ${\cal G}$ between equations~(\ref{eq2})
and (\ref{eq3}) and utilizing equation~(\ref{eq8}), we find that
the energy transport rate per unit mass is given by
\begeq
\epsilon \equiv {\dot E \over \dot M} = {1 \over 2} \, v^2
- {1 \over 2} \, {\ell^2 \over r^2} +  {\ell_0 \ell \over r^2}
+ {\gamma \over \gamma-1} \, a^2 + \pseudophi \ .
\label{eq15}
\fineq
Since we are working within the ADAF framework, $\epsilon$ will remain
constant unless a shock is formed in the disk. In this scenario, the
variation of the internal energy density $U$ is regulated by the
adiabatic compression of the gas and the viscous dissipation in the
disk. The comoving rate of change of $U$ can therefore be written in the
frame of the gas as (e.g., Becker \& Le 2003),
\begeq
{DU \over Dt} \equiv - v{dU \over dr} = -\gamma{U \over \rho}\,v{d\rho
\over dr} +{\rho \nu \over r^2} \left({d\ell \over dr} - {2\ell \over
r}\right)^2.
\label{eq16}
\fineq

We can obtain the dynamical differential equation governing the spatial
variation of the inflow velocity $v$ by manipulating the conservation
equations as follows. First we combine equations~(\ref{eq1}),
(\ref{eq13}), and (\ref{eq8}) to express the pressure as
\begeq
P = {\dot M a \ellK \over 4 \pi r^3 v} \ .
\label{eq17}
\fineq
Using this relation to substitute for the pressure in the radial
acceleration equation~(\ref{eq14}), we find that the derivative of the
sound speed $a$ is given by
\begeq
{da \over dr} = \left({a^2 \over v^2} - 1\right)
{v \over a} \, {dv \over dr} - {a \over \ellK} {d\ellK \over dr} 
+ {3 \, a \over r} + {\ell^2 - \ellK^2 \over a r^3}
\ .
\label{eq18}
\fineq
Next we differentiate equation~(\ref{eq15}) and combine the result with
equation~(\ref{eq11}) to obtain
\begeq
{dv \over dr} = {\ell^2-\ellK^2 \over v r^3} - {2 \gamma \over \gamma-1}
\, {a \over v} \, {da \over dr} - {\ellK (\ell - \ell_0)^2 \over \alpha
a^2 r^4}
\ .
\label{eq19}
\fineq
By using equation~(\ref{eq18}) to substitute for $da/dr$ in
equation~(\ref{eq19}), we obtain the differential dynamical equation
(e.g., Narayan et al. 1997)
\begeq
\left({v^2 \over a^2} - {2 \gamma \over \gamma+1} \right)
{1 \over v}{dv \over dr} = {2 \gamma \over \gamma+1}
\left({3 \over r} - {1 \over \ellK}{d\ellK \over dr}\right)
+\left({\gamma - 1 \over \gamma + 1}\right)
{v \ellK (\ell -\ell_0)^2 \over \alpha a^4 r^4} 
+ {\ell^2 - \ellK^2 \over a^2 r^3} \ . 
\label{eq20}
\fineq
In order to obtain the distributions of the flow variables $v$, $a$, and
$\ell$, we must solve numerically the governing equations~(\ref{eq11}),
(\ref{eq15}), and (\ref{eq20}). This requires consideration of the
critical structure of the flow as discussed below.

\subsection{Critical Conditions}

A black hole will swallow matter either from its binary companion or
from the winds of the surrounding stars. At the outer edge of the disk,
the accreting matter has negligible radial velocity, although it
ultimately crosses the event horizon with a local velocity equal to the
speed of light. Hence accretion flows around black holes must be
transonic. The locations where the flow transitions from subsonic to
supersonic are referred to as critical points. Depending on the
parameters describing the properties of the gas at a large distance from
the black hole, the flow may possess multiple critical points. When this
occurs, it is interesting to investigate the possible existence of
isothermal standing shock waves located between two of the critical
points. The location of the shock, if one exists, is determined by
applying the shock jump conditions. We first study the critical
structure of the transonic flow by rewriting equation~(\ref{eq20})
in the equivalent form
\begeq
{dv \over dr} = {N \over D} \ ,
\label{eq21}
\fineq
where the numerator and denominator functions $N$ and $D$ are defined by
\begeq
N \equiv {2 \gamma v \over \gamma+1}
\left({3 \over 2 r} + {1 \over r - \rs}\right)
+\left({\gamma - 1 \over \gamma + 1}\right)
{v^2 \ellK (\ell - \ell_0)^2 \over \alpha a^4 r^4} 
+ {v (\ell^2 - \ellK^2) \over a^2 r^3} \ ,
\label{eq22}
\fineq
and  
\begeq
D \equiv {v^2 \over a^2} - {2\gamma \over \gamma+1} \ .
\label{eq23}
\fineq

In order to obtain well-behaved global solutions, the numerator and
denominator functions must vanish at exactly the same location in the
flow. This condition represents the mathematical definition of the
critical point. Setting $N = D = 0$, we obtain the critical conditions
\begeq
{2 \gamma v_c \over \gamma+1}
\left({3 \over 2 r_c} + {1 \over r_c - \rs}\right)
+\left({\gamma - 1 \over \gamma + 1}\right)
\left[ {v_c^2 \ellK (\ell_c - \ell_0)^2 \over \alpha a_c^4 r_c^4}\right]
+ \left[{v_c (\ell_c^2 - \ellK^2) \over a_c^2 r_c^3}\right] =0 \ ,
\label{eq24}
\fineq
and
\begeq
{v^2_c \over a^2_c} - {2\gamma \over \gamma+1} \ = 0 \ ,
\label{eq25}
\fineq
where $r_c$, $v_c$, $a_c$, and $\ell_c$ denote the values of the radius,
the velocity, the sound speed, and the specific angular momentum at the
critical point, respectively. Any accretion solution must pass smoothly
through the critical point, which is guaranteed if the critical
conditions are satisfied. In this work, we assume approximate
equipartition between gas and magnetic pressure following Narayan et al.
(1997), and accordingly we set $\gamma = 1.5$ in the subsequent
analysis.

The topological nature of a critical point is determined by the value of
the velocity gradient $dv/dr$ at that point (e.g., Das 2007; Becker \&
Le 2003). In general, we obtain two possible values for $dv/dr$, with
one corresponding to accretion and the other to an outflow (wind). Any
physically acceptable solution must pass through a saddle type or X-type
critical point which is obtained when both of the derivatives are real
and of opposite signs. If a shock forms, then the accretion flow must
also pass through another saddle type critical point in the post-shock
region. When both of the values of $dv/dr$ are complex, the critical
point is referred to as O-type. This type of critical point is
unphysical and therefore no acceptable flow solution can pass through
such a point.

\subsection{Inner Boundary Conditions}

Solutions for the flow variables $v(r)$, $\ell(r)$, and $a(r)$ are
obtained by numerically integrating equations~(\ref{eq11}) and
(\ref{eq20}), supplemented by the algebraic relation given by
equation~(\ref{eq15}). Becker \& Le (2003) determined that in order to
ensure the stability of the calculation, the integration must proceed in
the outward direction, starting from a point close to the event horizon.
This requires the availability of asymptotic boundary conditions that
can be used to establish the starting values for the physical variables.
Based on the fundamental general relativistic principle that the torque
must vanish at the horizon (Weinberg 1972), Becker \& Le (2003) derived
the explicit asymptotic behaviors of $v(r)$, $\ell(r)$, and the
``entropy function,'' $K(r)$, for material subject to the
Shakura-Sunyaev viscosity prescription, where
\begeq
K(r) \equiv {r \, v \, a^{\gamma+1 \over \gamma-1} \over \Omega_{\rm K}}
\ .
\label{eq26}
\fineq
The physical significance of $K$ can be understood by combining
equations~(\ref{eq1}), (\ref{eq13}), (\ref{eq8}), and (\ref{eq26}) to
show that
\begeq
K^{\gamma-1} \propto {P \over \rho^\gamma} \ .
\label{eq27}
\fineq
This relation establishes that $K$ remains constant in regions of the
flow unaffected by dissipation, where $P \propto \rho^\gamma$.

If the gas is in local thermodynamic equilibrium, then one can
demonstrate that the value of $K$ is related to the entropy per particle
$S$ by (Reif 1965)
\begeq
S = k \, \ln K + c_0 \ ,
\label{eq28}
\fineq
where $c_0$ is a constant that depends only the composition of the gas,
but is independent of its state. Becker \& Le (2003) established that
the asymptotic behaviors of $K(r)$ and $\ell(r)$ close to the event
horizon are given by
\begeq
K(r) = K_0 \left[1 - {16 \, \alpha \, \ell_0^2 \over 5 \, c^2 \rs^5}
\left(\rs \over 2 \right)^{1/2} (r-\rs)^{5/2} \right] \ , \ \ \ \
r \to \rs \ ,
\label{eq29}
\fineq
and
\begeq
\ell(r) = \ell_0 +
{2 \, \alpha \, \ell_0 \over c^2 \rs} \left(2 \over \rs \right)^{1/2}
\left(K_0^2 \over 2 \, \rs^3\right)^{\gamma-1 \over \gamma+1}
(r - \rs)^{\gamma + 5 \over 2 \gamma + 2} \ , \ \ \ \ r \to \rs \ ,
\label{eq30}
\fineq
respectively, where $K_0$ represents the entropy value at the horizon.
Note that the asymptotic radial dependence of $K(r)$ is very weak,
reflecting the unimportance of viscous dissipation near the horizon.
Becker \& Le (2003) also found that the asymptotic variation of the
inflow velocity is given by
\begeq
v(r) = \vff(r) \left[1 +
{2 \, \epsilon_0 r^2 - \ell_0^2 - (\gamma+1) \, f(r)
\over r^2 \, \vff^2(r) - (\gamma-1) \, f(r)} \right]^{1/2}
 \ , \ \ \ \
r \to \rs \ ,
\label{eq31}
\fineq
where $\epsilon_0$ denotes the value of the specific energy transport
rate at the horizon, the free-fall velocity is given by
\begeq
\vff(r) = \left(2 \, GM \over r - \rs\right)^{1/2} \ ,
\label{eq32}
\fineq
and the function $f$ is defined by
\begeq
f(r) \equiv {2 \, \gamma \, r^2 \over \gamma^2-1} \left[K_0^2 \over 2 \,
r^3 (r-\rs) \right]^{\gamma-1 \over \gamma+1} \ .
\label{eq33}
\fineq
For given values of the parameters $\epsilon_0$, $\ell_0$, and $K_0$,
equations~(\ref{eq30}) and (\ref{eq31}) can be used to compute starting
values for the integration of the variables $\ell(r)$ and $v(r)$ away
from the event horizon.

\section{GLOBAL SOLUTIONS}

The structure of the global disk/shock model depends on the energy
transport rate at the horizon $\epsilon_0$, the angular momentum at the
horizon $\ell_0$, the entropy at the horizon $K_0$, the viscosity
parameter $\alpha$, and the ratio of specific heats $\gamma$. In the
calculations performed here, we set $\gamma = 1.5$, following Narayan et
al. (1997). When a shock is present, we use the subscripts ``-'' and
``+'' to denote quantities measured just upstream and just downstream
from the shock, respectively. In a shocked disk, the gas passes through
one critical point in the pre-shock region at $r=r_c^{\rm out}$, and
then through another in the post-shock region at $r=r_c^{\rm in}$
(Abramowicz \& Chakrabarti 1990). Conversely, if the flow is smooth
(shock-free), then the gas only passes through one critical point,
located at radius $r=r_c$.

The process of determining the structure of a viscous ADAF disk
containing an isothermal shock begins with the selection of input values
for $\epsilon_0$, $\ell_0$, $\alpha$, and $\gamma$. We must also select
a provisional value for the entropy parameter $K_0$. Based on this
information, we can utilize the asymptotic relations given by
equations~(\ref{eq30}) and (\ref{eq31}) to compute values for $\ell(r)$
and $v(r)$, respectively, at the starting radius for the outward
integration, which is taken to be $2.001\,GM/c^2$. Next we numerically
integrate equations~(\ref{eq11}), (\ref{eq15}), and (\ref{eq20}),
beginning at the starting radius and working outward towards the inner
critical point. The value of $K_0$ is varied until a solution is
obtained that satisfies the critical conditions given by
equations~(\ref{eq24}) and (\ref{eq25}), which ensure that the flow
passes smoothly through the inner critical point at radius $r_c^{\rm
in}$.

After the location of the inner critical point has been established, the
integration must be continued into the upstream, subsonic region. Due to
the mathematical structure of the dynamical equation~(\ref{eq21}), it is
not possible to integrate across the critical point itself. We must
therefore employ linear extrapolation, which requires the calculation of
the derivative $dv/dr$ at the critical point using l'H\^opital's rule.
After performing the linear extrapolation, the integration is continued
in the subsonic region until we reach the shock radius, $r_*$, which is
assigned a provisional value initially. Since no energy is lost from the
disk between the shock and the horizon, it follows that $\epsilon_0 =
\epsilon_+$, where $\epsilon_+$ is the energy transport rate on the
downstream side of the shock. At the shock location, the velocity and
the energy transport rate experience discontinuous transitions described
by the isothermal shock jump conditions, which can be written as (e.g.,
Chakrabarti 1989a)
\begeq
{v_+ \over v_-} = {\cal M}_+^2
\ , \ \ \ \ \
\Delta\epsilon \equiv \epsilon_+-\epsilon_- = {v_+^2 - v_-^2 \over 2}
\ ,
\label{eq34}
\fineq
where ${\cal M}_+ \equiv v_+/a_+$ denotes the post-shock Mach number.
This relation facilitates the calculation of the pre-shock (incident)
energy transport rate, $\epsilon_-$. Note that if no shock is present,
then $\epsilon$ is continuous throughout the entire flow, and therefore
$\Delta\epsilon=0$ and $\epsilon_0 = \epsilon_-$. By definition, the
flow maintains a uniform temperature across the isothermal shock, and
consequently $a_+ = a_-$. The shock radius $r_*$ is varied until we
obtain a solution that passes smoothly through the outer critical point,
located in the upstream region at radius $r_c^{\rm out}$. The solution
process is completed when the critical conditions are satisfied at the
outer critical point. For given input values of $\epsilon_0$, $\ell_0$,
$\alpha$, and $\gamma$, in the end we obtain unique values for $K_0$,
$r_*$, $r_c^{\rm in}$, and $r_c^{\rm out}$, in addition to the
associated global dynamical solution for the disk/shock structure.

\begin{figure}
\centering
\includegraphics[scale=0.5]{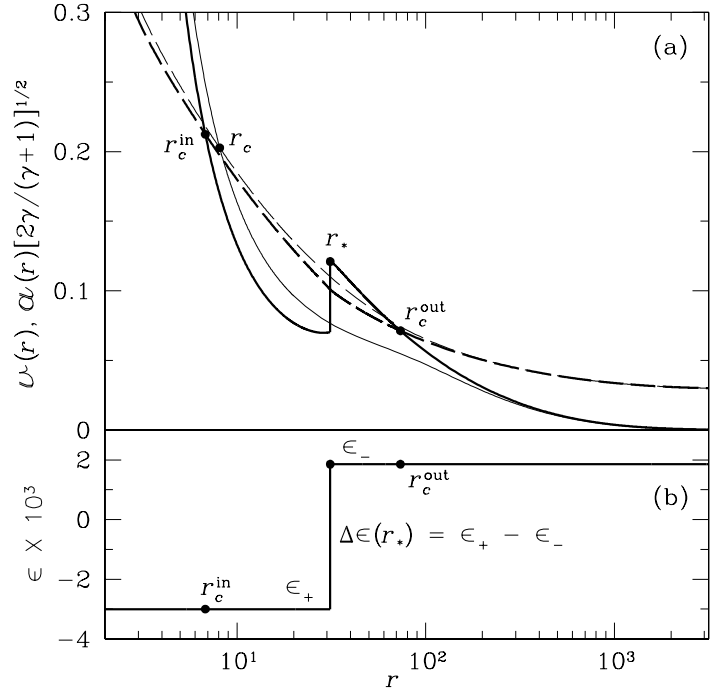}
\caption{({\it a}) Inflow velocity $v$ (solid lines) and isothermal sound
speed $a$ multiplied by $\sqrt{2\gamma/(\gamma+1)}$ (dashed lines)
plotted as functions of the radius $r$ in units of $GM/c^2$ for a
typical global accretion solution with $\gamma=1.5$. The thick and thin
lines represent the shocked and smooth solutions, respectively. ({\it b})
Energy transport rate $\epsilon$ plotted as a function of radius $r$.
Note the jump to the negative post-shock value due to the release of
energy through the upper and lower surfaces of the accretion disk at the
isothermal shock location. See the text for details.}
\end{figure}

\begin{figure}
\centering
\includegraphics[scale=0.8]{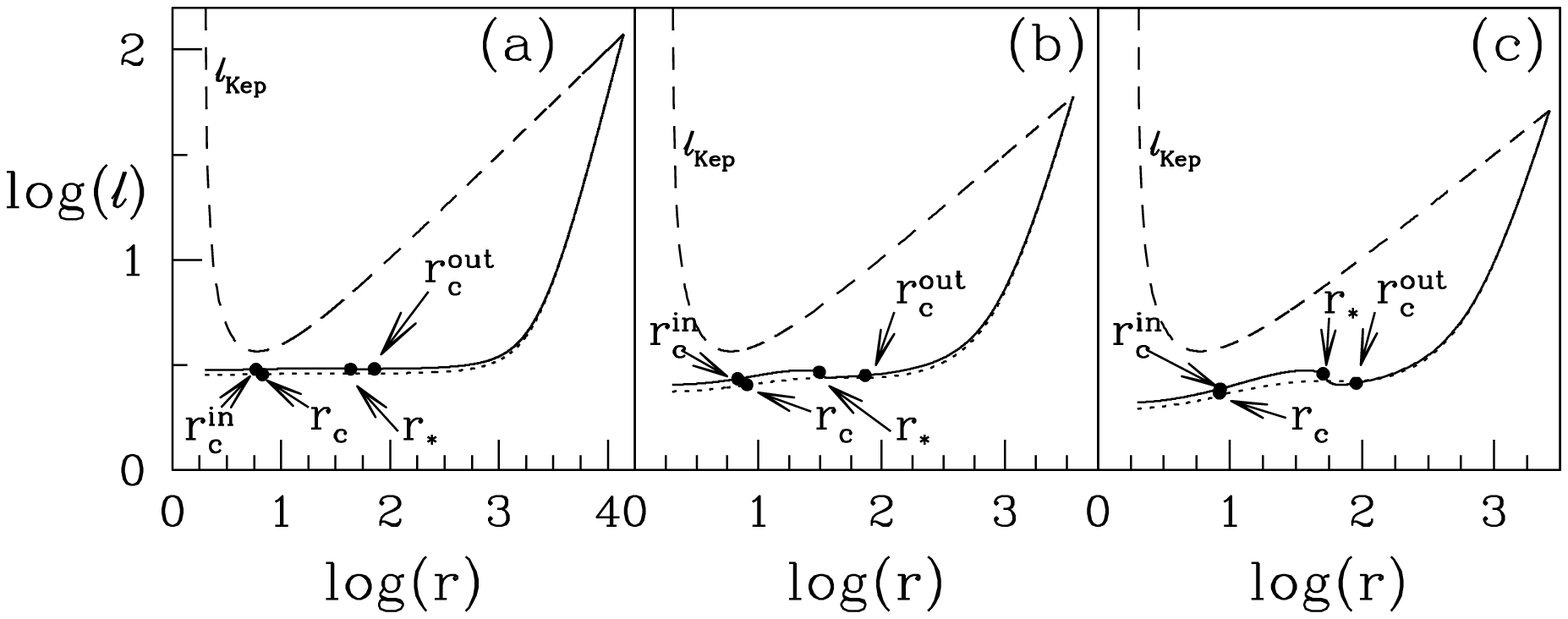}
\vskip-1.5truein
\caption{Specific angular momentum $\ell$ plotted as a function of the
radius $r$ for ({\it a}) $\alpha = 0.01$, ({\it b}) $\alpha = 0.1$, and
({\it c}) $\alpha = 0.2$. Here we assume that $\gamma=1.5$. The solid,
dotted, and dashed lines correspond to the shocked, smooth, and
Keplerian profiles, respectively. See the text for the complete sets of
parameters used in each panel.}
\end{figure}

A typical set of shocked and smooth accretion solutions obtained for the
same value of the incident energy transport rate $\epsilon_-$ is
presented in Figure~1{\it a}. In these calculations, the Shakura-Sunyaev
viscosity parameter was specified by setting $\alpha = 0.1$, and the
specific heats ratio is given by $\gamma=1.5$. By applying the
multi-step iterative method described above, we obtain for the shocked
flow the parameters $\epsilon_- = 0.001856$, $\epsilon_0 = \epsilon_+ =
-0.003$, $\ell_0 = 2.55$, and $K_0 = 0.00689$. In the shocked case, the
incident flow first becomes supersonic upon passing through the outer
critical point located at $r_c^{\rm out} = 73.51$. The gas subsequently
passes through an isothermal shock at $r_* = 31.2$, where the velocity
makes a discontinuous transition to become subsonic again. Downstream
from the shock, the flow accelerates towards the horizon, and regains
its supersonic character after crossing the inner critical point at
$r_c^{\rm in} = 6.798$. The radius at the outer edge of the hot region,
$r_{\rm edge} = 3555$, is computed by setting $\ell=\ellK$, so that the
angular momentum matches the local Keplerian value. The radial profile
of the specific angular momentum $\ell$ is plotted in Figure~2{\it b}.
Although the incident energy transport rate $\epsilon_-$ is negative in
our calculations, it should be emphasized that the total energy inflow
rate, including the rest mass energy, is positive as required.

For the smooth-flow solution depicted in Figure~1{\it a}, we used the
parameters $\epsilon_0 = \epsilon_+ = \epsilon_- = 0.001856$, $\ell_0 =
2.368$, and $K_0 = 0.008565$. In this case, the value of $\ell_0$ was
selected so that $\ell=\ellK$ at the same outer edge radius, $r_{\rm
edge} = 3555$, obtained in the shocked disk (see Fig.~2{\it b}). The
smooth flow solution passes through a single critical point, located at
$r_c=8.097$. It is interesting to compare the plots of the sound speed
obtained in the smooth and shocked cases. The loss of energy at the
shock significantly reduces the sound speed (and therefore the
temperature) in the post-shock region relative to the smooth solution.
In Figure~1{\it b}, we plot the variation of the energy transport rate
for the shock solution. The excess energy released through the upper and
lower surfaces of the disk at the shock location may be sufficient to
power the formation of a relativistic jet (Le \& Becker 2005; Becker,
Das, \& Le 2008).

The viscosity parameter $\alpha$ plays a central role in determining the
efficiency of the angular momentum transport in the disk. We explore the
effect of varying this parameter in Figure~2, where we plot the results
obtained for the radial profile of the specific angular momentum $\ell$
in both shocked and smooth disks for three different values of $\alpha$.
All of the solutions are sub-Keplerian ($\ell < \ellK$) for $r < r_{\rm
edge}$. In the shocked case, the increase in the angular momentum in the
post-shock region produces the ``centrifugal barrier'' required for
shock formation, while no such barrier exists in the smooth case. In
each of the panels, the value of $\ell_0$ was determined by satisfying
the inner critical conditions, and the incident energy transport rate
$\epsilon_-$ has the same value in the shocked and smooth scenarios. The
parameter values adopted in panel ({\it a}) are $\alpha = 0.01$, $\gamma
= 1.5$, $\epsilon_+ = -1.0 \times 10^{-3}$, $\epsilon_- = 1.9072 \times
10^{-3}$, $\ell_0 = 3.0$, $K_0 = 2.7509 \times 10^{-3}$ in the shocked
case, and $\epsilon_+ = \epsilon_- = 1.9072 \times 10^{-3}$, $\ell_0 =
2.845$, $K_0 = 3.8588 \times 10^{-3}$ in the smooth (shock-free) case. In
panel ({\it b}), we adopt the same parameters used in Figure~1, namely
$\alpha = 0.1$, $\gamma = 1.5$, $\epsilon_+ = -3.0 \times 10^{-3}$,
$\epsilon_- = 1.856 \times 10^{-3}$, $\ell_0 = 2.55$, $K_0 = 6.89 \times
10^{-3}$ in the shocked case, and $\epsilon_+ = \epsilon_- = 1.856
\times 10^{-3}$, $\ell_0 = 2.368$, $K_0 = 8.565 \times 10^{-3}$ in the
smooth case. The values adopted in panel ({\it c}) are $\alpha = 0.2$,
$\gamma = 1.5$, $\epsilon_+ = -1.0 \times 10^{-3}$, $\epsilon_- = 1.6483
\times 10^{-3}$, $\ell_0 = 2.1$, $K_0 = 1.1883 \times 10^{-2}$ in the
shocked case, and $\epsilon_+ = \epsilon_- = 1.6483 \times 10^{-3}$,
$\ell_0 = 1.966$, $K_0 = 1.25 \times 10^{-2}$ in the smooth case. Note
the increasing prominence of the centrifugal barrier as $\alpha$
increases.

\section{SHOCK PROPERTIES}

Since this is the first work to extensively consider the consequences
of a standing shock for the structure of a viscous accretion disk,
it is interesting to examine how the shock properties
depend on the various flow parameters. We carry out the analysis for
flows characterized by the fundamental parameters ($\epsilon_0$, $\ell_0$,
$\alpha$), where $\epsilon_0$ denotes the energy transport rate at the
horizon, which is the same as the post-shock energy transport rate
$\epsilon_+$. In each case, we keep one of the three parameters
($\epsilon_0$, $\ell_0$, $\alpha$) fixed and vary the other two to study
the resulting shock properties.

\begin{figure}
\centering
\includegraphics[scale=0.5]{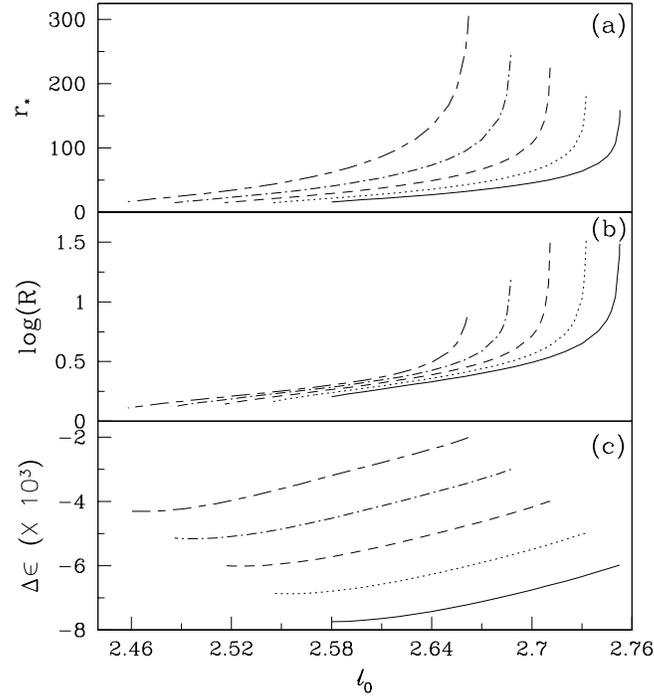}
\caption{Variation of ({\it a}) the isothermal shock radius $r_*$, ({\it
b}) the compression ratio $R = v_-/v_+$, and ({\it c}) the energy jump
$\Delta \epsilon$ as functions of $\ell_0$. The solid, dotted, dashed,
dot-dashed, and short-long-dashed curves correspond to $\epsilon_0 =
-0.006$, $-0.005$, $-0.004$, $-0.003$, and $-0.002$, respectively. Here,
we set $\alpha = 0.1$ and $\gamma=1.5$.}
\end{figure}

In Figure~3 we present the results obtained for the shock radius, $r_*$,
the compression ratio, $R = v_-/v_+$, and the energy jump,
$\Delta\epsilon = \epsilon_+-\epsilon_-$, as functions of the angular
momentum transport rate $\ell_0$ and the energy transport rate at the
horizon $\epsilon_0$. In these calculations, the viscosity and specific
heat parameters are held constant, with $\alpha = 0.1$ and $\gamma=1.5$,
respectively. Figure~3{\it a} clearly demonstrates that accretion flows
with a very wide range of input parameters may possess isothermal
standing shocks. Note that for a given post-shock energy parameter
$\epsilon_0$, the shock location recedes away from the black hole with
increasing $\ell_0$ because of the enhancement of the centrifugal
barrier. On the other hand, the shock location shifts inward as
$\epsilon_0$ decreases for a given value of $\ell_0$, due to the
increased ram pressure of the flow. For a given value of $\epsilon_0$,
there is a restricted range of $\ell_0$ values within which a shock can
exist. In general, we find that the shock radius $r_*$ always exceeds
$\sim 10\,GM/c^2$. In Figure~3{\it b}, we plot the variation of the
shock compression ratio $R$ as a function of the flow parameters
$\epsilon_0$ and $\ell_0$. We observe that for fixed $\epsilon_0$, the
compression ratio increases as the shock moves away from the black hole.
In Figure~3{\it c}, we plot the corresponding variation of the energy
jump $\Delta\epsilon$. Note that $\Delta\epsilon$ becomes more negative
as $\epsilon_0$ decreases for fixed $\ell_0$, which reflects the fact
that a larger amount of gravitational potential energy is released into
the outflow when the shock wave forms closer to the black hole.

\begin{figure}
\centering
\includegraphics[scale=0.5]{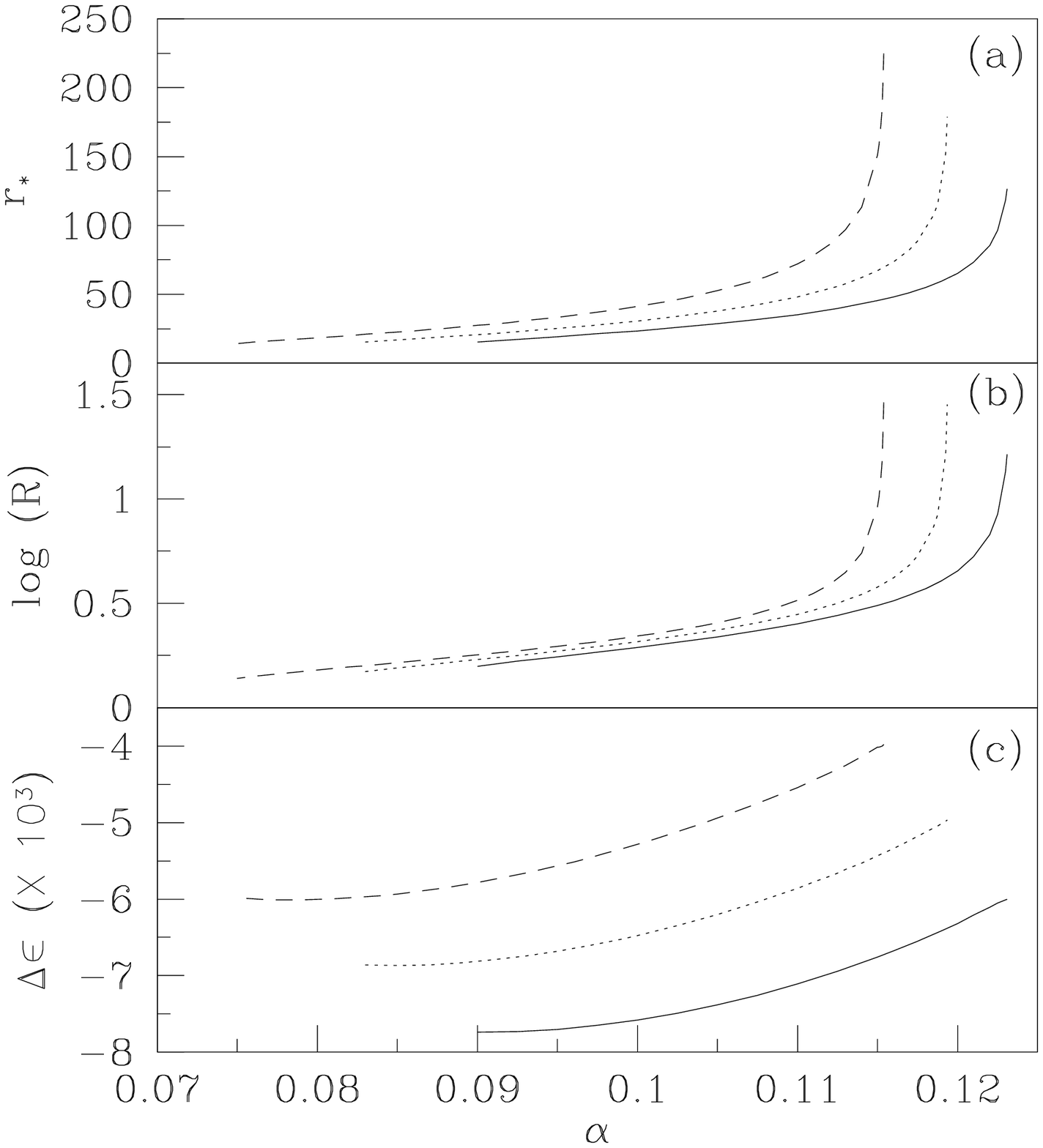}
\caption{Variation of ({\it a}) the isothermal shock radius $r_*$, ({\it
b}) the compression ratio $R = v_-/v_+$, and ({\it c}) the energy jump
$\Delta \epsilon$ as functions of $\alpha$. The solid, dotted, and
dashed lines denote $\epsilon_0 = -0.006$, $-0.005$, and $-0.004$,
respectively. Here, we set $\ell_0 = 2.62$ and $\gamma=1.5$.}
\end{figure}

Next we examine the dependence of the shock properties on the viscosity
parameter $\alpha$. In Figure~4{\it a}, the isothermal shock location
$r_*$ is plotted as a function of $\alpha$ for various values of the
post-shock (horizon) energy transport rate $\epsilon_0$ for the case
with $\gamma=1.5$. Here, we fix the angular momentum transport rate by
setting $\ell_0 = 2.62$. In all cases, we observe that shocks can form
over a wide range of flow parameters. For a given $\alpha$, the shock
tends to form farther away from the black hole as $\epsilon_0$ is
increased, which is consistent with the behavior noted in Figure~3{\it
a}. We also observe that for fixed $\epsilon_0$, the shock radius $r_*$
increases with $\alpha$. This is because higher values of $\alpha$ tend
to reduce the angular momentum, which forces the centrifugal barrier to
operate at a larger radius in order to be effective. Note that for a
given set of flow parameters ($\epsilon_0$, $\ell_0$), the viscosity
parameter $\alpha$ has both upper and lower limits for shock formation.
The lower limit reflects the fact that fully inviscid flows cannot form
shocks for arbitrary combinations of $\epsilon_0$ and $\ell_0$ (e.g., Le
\& Becker 2005). Conversely, the upper limit on $\alpha$ is due to the
efficient dissipation of angular momentum that occurs as $\alpha$
increases. This effect prevents the development of the centrifugal
barrier required for shock formation if $\alpha$ is too large.

\begin{figure}
\centering
\includegraphics[scale=0.5]{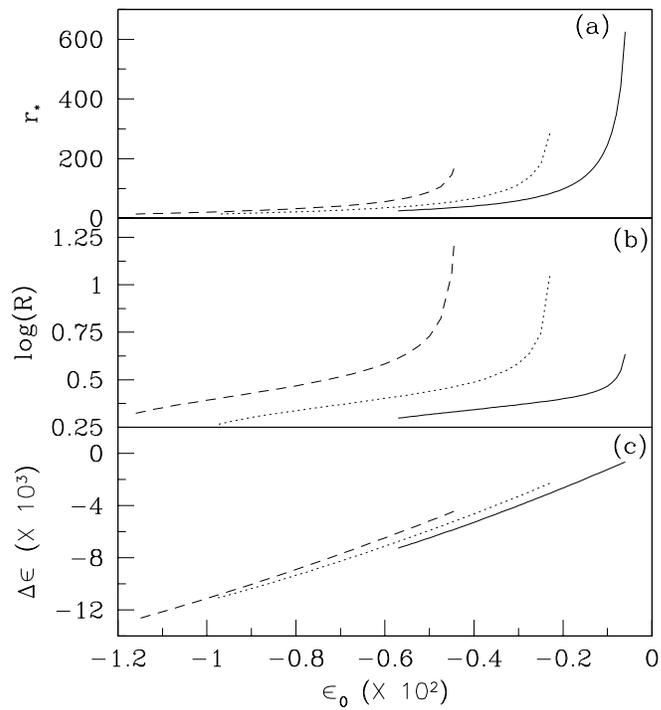}
\caption{Variation of ({\it a}) the isothermal shock radius $r_*$, ({\it
b}) the compression ratio $R = v_-/v_+$, and ({\it c}) the energy jump
$\Delta \epsilon$ as functions of $\epsilon_0$. The solid, dotted, and
dashed curves represent $\ell_0 = 2.62$, 2.67, and 2.72, respectively.
Here, we set $\alpha = 0.1$ and $\gamma=1.5$.}
\end{figure}

The dependence of the shock properties on post-shock energy $\epsilon_0$
is illustrated in Figure~5 for three values of the angular momentum
transport rate $\ell_0$. The viscosity and specific heat parameters are
held constant with the values $\alpha = 0.1$ and $\gamma=1.5$,
respectively. In Figure~5{\it a} we plot the shock location $r_*$ as a
function of $\epsilon_0$. Since the standing shock is centrifugally
supported, the shock radius $r_*$ increases when $\ell_0$ is increased
for a fixed value of $\epsilon_0$. On the other hand, if we hold $r_*$
constant, then we observe an anticorrelation between $\ell_0$ and
$\epsilon_0$. In Figure~5{\it b} and Figure~5{\it c}, respectively, we
depict the associated variations of the compression ratio $R$ and the
energy jump $\Delta \epsilon$ as functions of $\epsilon_0$ for the same
set of flow parameters used in Figure~5{\it a}. The compression ratio
increases when either $\ell_0$ or $\epsilon_0$ is increased. We note
that $\Delta \epsilon$ increases almost linearly with increasing
$\epsilon_0$ for fixed $\ell_0$. The results indicate that flows with
larger values of $\ell_0$ tend to discharge less energy at the shock
location.

\begin{figure}
\centering
\includegraphics[scale=0.5]{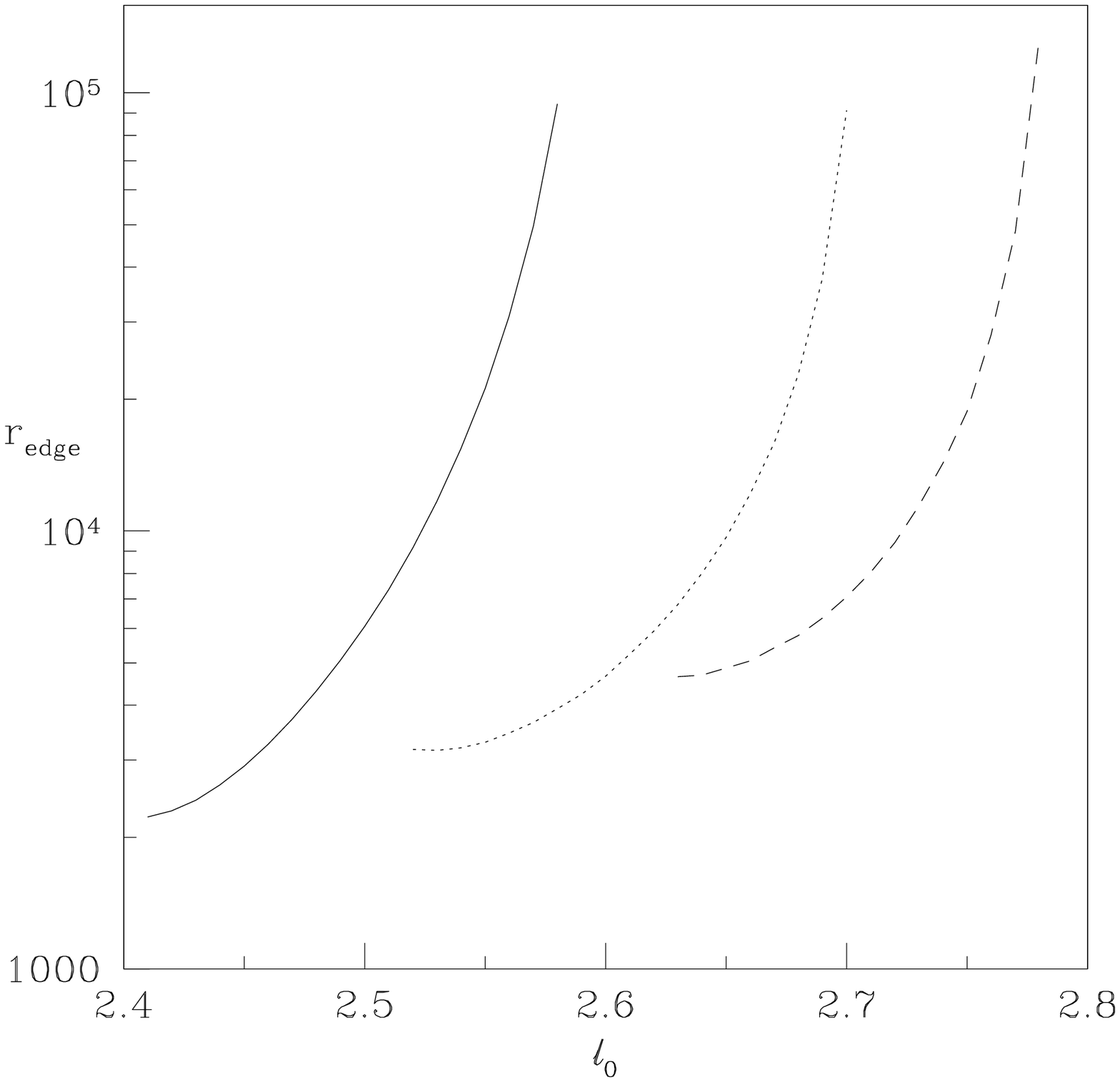}
\caption{Plot of the outer edge radius $r_{\rm edge}$ as a function of
$\ell_0$. This is the radius where the angular momentum $\ell$ equals
the Keplerian value $\ellK$ (eq.~[\ref{eq12}]). The solid, dotted,
and dashed lines correspond to $\epsilon_0 = 0.0$, $-0.004$, and $-0.008$,
respectively. Here, we set $\alpha = 0.1$ and $\gamma=1.5$. See the text
for details.}
\end{figure}

The hot, sub-Keplerian ADAF region must connect with a cool Keplerian
outer disk at some large radius. We estimate the outer edge radius of
the ADAF zone, $r_{\rm edge}$, by setting the angular momentum of the
accretion flow equal to the local Keplerian value. In Figure~6 we plot the
variation of the outer edge radius as a function of $\ell_0$ and
$\epsilon_0$ for the case with $\alpha = 0.1$ and $\gamma=1.5$. We
observe that for a fixed value of $\epsilon_0$, the value of $r_{\rm
edge}$ rapidly increases with increasing $\ell_0$, indicating that the
cool Keplerian disk recedes from the horizon. The outer edge radius also
increases with increasing $\epsilon_0$ when $\ell_0$ is held fixed.

\begin{figure}
\centering
\includegraphics[scale=0.6]{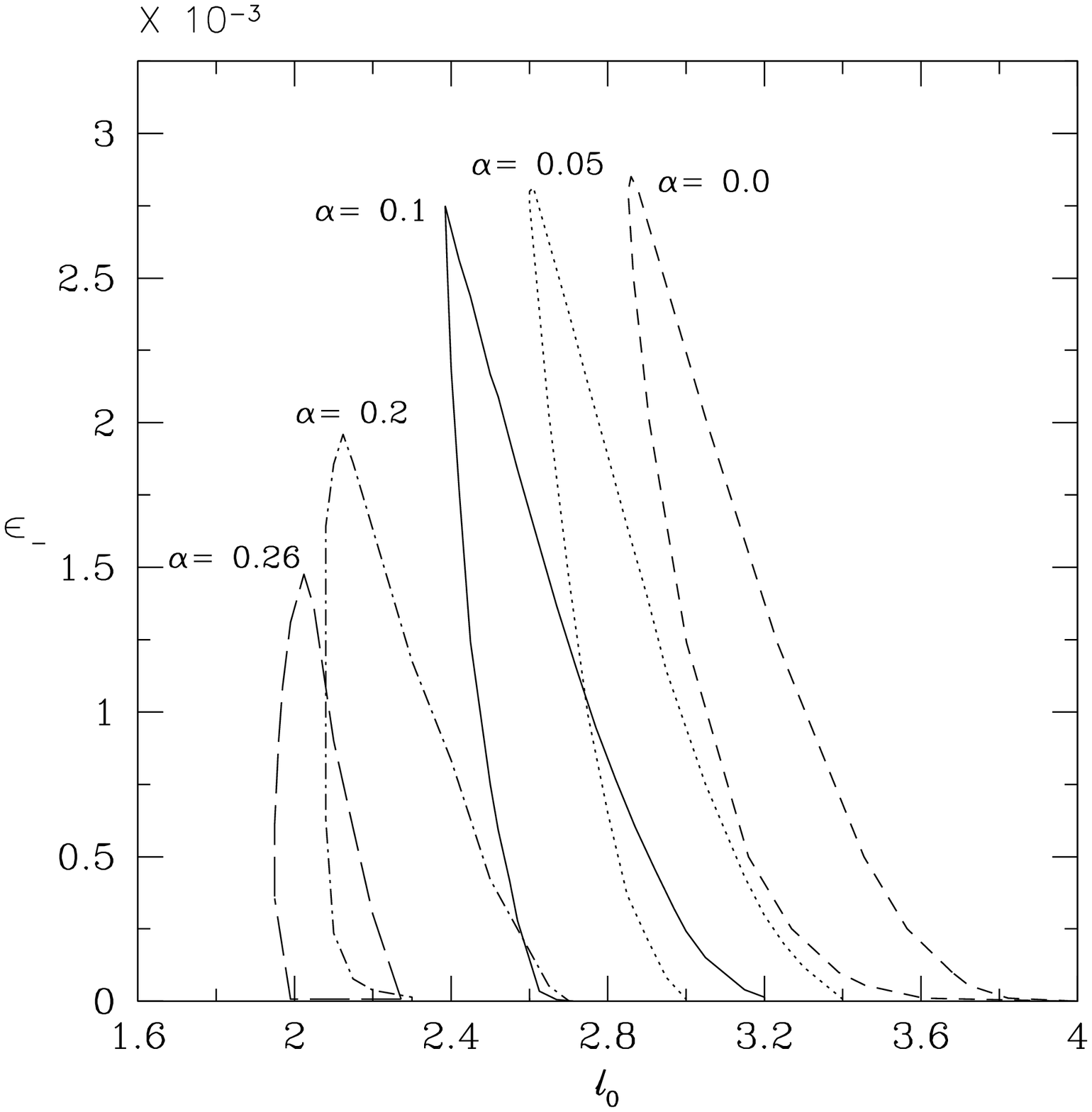}
\caption{Representation of the parameter space spanned by the specific
angular momentum at the horizon $\ell_0$ and the incident energy
transport rate $\epsilon_-$. The wedge-shaped regions denote the areas
of the parameter space within which shocks can form for the indicated
values of $\alpha$. See the discussion in the text.}
\end{figure}

The shocked, viscous, advection-dominated disks discussed here represent
a new type of astrophysical phenomenon, and therefore it is important to
investigate the range of flow parameters within which shocks can form.
In Figure~7 we analyze this question for the case with $\gamma=1.5$ by
varying the angular momentum transport rate $\ell_0$ and the upstream
(incident) energy transport rate $\epsilon_-$ while holding the
viscosity parameter $\alpha$ fixed at five different values. The solid
curve depicts the region within which shocked solutions are possible for
fully inviscid flow ($\alpha=0$). This result agrees very well with
Figure~2 from Le \& Becker (2005). We note that the allowed region for the
shocked solution shifts toward lower angular momentum as $\alpha$
increases. This shift occurs because the specific angular momentum of
the accreting material at the horizon, $\ell_0$, decreases in response
to the enhanced viscosity associated with the increase in $\alpha$. The
shrinkage of the region for the shocked solution observed as $\alpha$
increases suggests that shocked disk solutions are impossible above a
critical value of $\alpha$, denoted by $\alpha_{\rm crit}$. For the
model considered here, with $\gamma=1.5$, we find that $\alpha_{\rm
crit} \approx 0.27$.

\section{ASTROPHYSICAL APPLICATIONS}

It is interesting to ask whether the self-consistent disk/shock model we
have described can account for the energetics of specific jet sources,
such as M87 and \SgrA. When observational values of the black hole mass
$M$, the accretion rate $\dot M$, and the jet kinetic luminosity $\Qjet$
are available for a particular source, we can utilize this information
to constrain the model parameters. The process begins by selecting
values for the fundamental free parameters $\epsilon_0$, $\alpha$, and
$\gamma$. After selecting provisional values for $\ell_0$ and $K_0$, we
utilize the asymptotic relations given by equations~(\ref{eq30}) and
(\ref{eq31}) to compute initial values for the functions $\ell(r)$ and
$v(r)$ at the starting radius $2.001\,GM/c^2$. We then numerically
integrate equations~(\ref{eq11}), (\ref{eq15}), and (\ref{eq20}) in the
outward direction, and vary the value of $K_0$ until the critical
conditions given by equations~(\ref{eq24}) and (\ref{eq25}) are
satisfied at the inner critical point.

The energy transport rate $\epsilon$ drops from its upstream value
$\epsilon_-$ to the downstream value $\epsilon_+ = \epsilon_0$ at the
shock location due to the escape of energy from the disk into the
outflow (jet). We can therefore determine the shock radius $r_*$ by
requiring that the value of $\Delta\epsilon$ computed using
equation~(\ref{eq34}) agrees with that calculated based on the observed
jet kinetic luminosity, $\Qjet$, using the relation
\begeq
\Qjet = - \dot M \Delta\epsilon
\ ,
\label{eq35}
\fineq
where the negative sign appears because $\Delta\epsilon < 0$. Once the
shock radius is established, the integration is continued on the
upstream side of the shock toward the outer critical point. If the flow
does not pass smoothly through the outer critical point, then the value
of $\ell_0$ is modified and the procedure is repeated starting from the
horizon. The computation of the self-consistent disk/shock dynamical
model is complete when the critical conditions are satisfied at the
outer critical point. For given input values of $\epsilon_0$, $\alpha$,
and $\gamma$, in the end we obtain the complete global dynamical
solution along with unique values for $\ell_0$, $K_0$, $r_*$, $r_c^{\rm
in}$, and $r_c^{\rm out}$. We provide two specific examples below.

\subsection{M87}

The disk/shock/outflow model we have developed was previously utilized
by Becker, Das, \& Le (2008) to analyze the structure of the accretion
flow in M87. The results obtained for that source are presented in more
detail here. Our application to M87 is based on the observational values
$\dot M = 1.34 \times 10^{-1} {\rm \ M_{\odot} \ yr^{-1}}$ (Reynolds et
al. 1996), $\Qjet = 5.5 \times 10^{43}\,{\rm erg \ s^{-1}}$ (Reynolds et
al. 1996; Bicknell \& Begelman 1996; Owen, Eilek, \& Kassim 2000), and
$M = 3.0 \times 10^9 {\rm \ M_{\odot}}$ (e.g., Ford et al. 1994). We set
the viscosity parameter $\alpha=0.1$ in order to demonstrate that shocks
can exist in ADAF disks even in the presence of substantial viscosity,
although our model can accommodate any value for $\alpha$. Based on the
assumption of approximate energy equipartition between the magnetic
field and the gas internal energy, we adopt the value $\gamma=1.5$ for
the adiabatic index (Narayan et al. 1997). We utilize natural
gravitational units in our numerical examples, with $GM=c=1$ and
$\rs=2$, except as noted. The remaining parameters for the shocked-disk
model can be computed based on the observations of M87, from which we
obtain $\epsilon_-=0.001516$, $\epsilon_+ = \epsilon_0 = -0.005746$,
$\ell_0=2.6257$, $K_0=0.00608$, $r_*=26.329$, $r_c^{\rm in}=6.462$, and
$r_c^{\rm out}=96.798$. For the pre-shock and post-shock velocities and
Mach numbers we obtain $u_-=0.138$, $u_+=0.068$, ${\cal M}_-=1.427$, and
${\cal M}_+=0.701$, respectively, and for the disk half-thickness at the
shock location we obtain $H_*=12.10$.

The results obtained for the inflow speed $u$ and the isothermal sound
speed $a$ are plotted in Figure~8{\it a}, and the associated solution
for the specific angular momentum $\ell$ is depicted along with the
Keplerian profile $\ellK$ (eq.~[\ref{eq12}]) in Figure~8{\it b}. Results
are presented for both shocked and smooth (shock-free) disks. In the
shocked disk, the source is located at the shock radius, $r_*=26.329$,
and we observe that the values of $\ell$ and $\ellK$ merge at $r_{\rm
edge}=4658$, which represents the outer edge of the ADAF region. The
smooth solution is computed by integrating the conservation equations
using the same value for the incident energy transport rate $\epsilon_-$
as adopted in the shocked-disk model, but with no shock included. The
value of $\ell_0$ is then varied until we obtain $\ell=\ellK$ at the
same outer radius, $r_{\rm edge}=4658$, as in the shocked disk. The
resulting parameter values for the smooth solution are $\epsilon_- =
\epsilon_+ = 0.001516$, $\ell_0=2.3988$, $K_0=0.0084$, and $r_c=7.572$.
In order to facilitate comparison with the shocked model, in the smooth
case the source is assumed to be at the same radius $r = 26.329$.

Our solutions for the specific angular momentum $\ell$ are
significantly sub-Keplerian, in agreement with the results obtained by
Narayan et al. (1997), who utilized the same set of ADAF conservation
equations employed here. The sub-Keplerian nature of the flow stems from
the relatively high value for the viscosity parameter adopted here,
$\alpha = 0.1$. Note that the sound speed (and hence the temperature) is
significantly lower in the shocked case as compared with the smooth disk
due to the release of energy into the outflow at the shock location.

It is important to examine the energy balance in the shocked model to
ensure that the rate of energy loss at the shock location is equal to
the observational value for the kinetic luminosity in M87. We can
compute the theoretical value for the jet luminosity, $\Qjet$, by
substituting our values for $\epsilon_+$ and $\epsilon_-$ into
equation~(\ref{eq35}). The result obtained is $\Qjet = 5.5 \times
10^{43}\,{\rm erg \ s^{-1}}$, which agrees with the observed jet
luminosity for M87. This confirms that our viscous shocked disk model is
able to account for the energetics of the M87 outflow.

\begin{figure}
\centering
\includegraphics[scale=0.7]{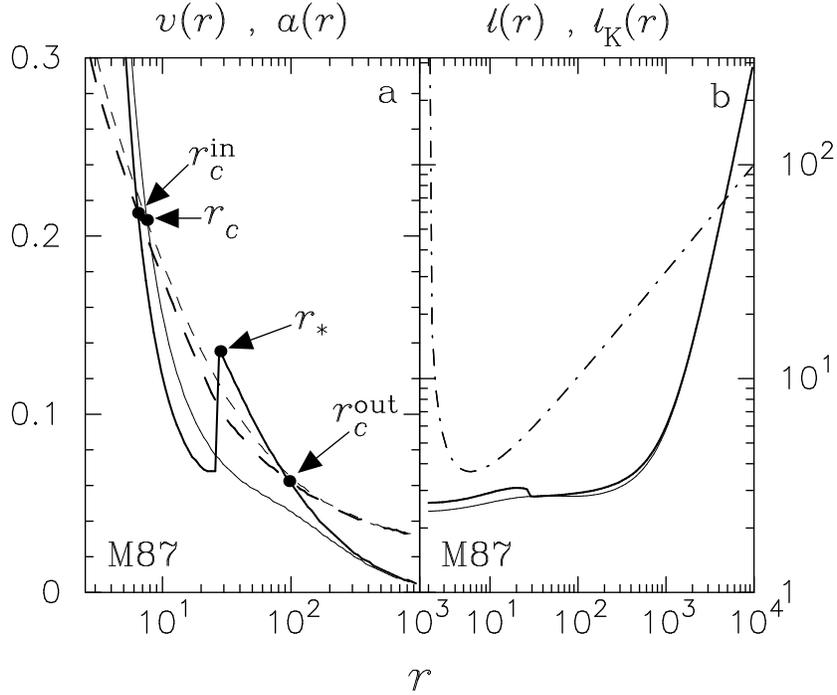}
\caption{Dynamical profiles for M87. ({\it a}) Inflow velocity $v$
(solid lines) and isothermal sound speed $a$ multiplied by
$\sqrt{2\gamma/(\gamma+1)}$ (dashed lines) plotted as functions of the
radius $r$ in units of $GM/c^2$. The thick and thin lines denote the
shocked and smooth solutions, respectively. ({\it b}) Specific angular
momentum $\ell$ for the shocked (thick line) and smooth (thin line)
solutions plotted as functions of $r$ in units of $GM/c^2$ along with
the Keplerian angular momentum $\ellK$ (dot-dashed line).}
\end{figure}

\subsection{\SgrA}

For \SgrA, we adopt the values $M = 2.6 \times 10^6 {\rm \ M_\odot}$
(Sch\"odel et al. 2002) and $\dot M = 8.96 \times 10^{-7} {\rm \ M_\odot
\ yr^{-1}}$ (Yuan, Markoff, \& Falcke 2002; Quataert 2003), and we set
$\gamma=1.5$ and $\alpha=0.1$. The kinetic luminosity of the jet is
estimated using $\Qjet = 5 \times 10^{38} {\rm ergs \ s^{-1}}$ (Falcke
\& Biermann 1999), although this value is rather uncertain (e.g., Yuan
2000; Yuan et al. 2002). The results for the remaining model parameters
implied by the observations are $\epsilon_-=0.00134884$, $\epsilon_+ =
\epsilon_0 = -0.0085$, $\ell_0=2.6728$, $K_0=0.005448$, $r_*=19.917$,
$r_c^{\rm in}=6.380$, and $r_c^{\rm out}=112.384$. For the pre-shock and
post-shock velocities and Mach numbers we obtain $u_-=0.159$,
$u_+=0.0748$, ${\cal M}_-=1.4578$, and ${\cal M}_+=0.6860$,
respectively, and for the disk half-thickness at the shock location we
obtain $H_*=8.72$.

In Figure~9{\it a} we plot the results obtained for the inflow speed
$u$ and the isothermal sound speed $a$ in both the shocked and smooth
disk models for \SgrA. The profiles of $\ell$ and $\ellK$ are plotted in
Figure~9{\it b}. In the shocked case, the source is located at
$r_*=19.917$, and we find that $\ell = \ellK$ at radius $r_{\rm
edge}=5432$, which is the outer edge of the ADAF region. The incident
energy transport rate $\epsilon_-=0.00134884$ is the same for both the
smooth and shocked models, and the value of $\ell_0$ in the smooth case
is determined by requiring that $\ell$ merge with the Keplerian profile
at the same radius $r_{\rm edge}=5432$ as in the shocked case. Based on
this approach, the parameters obtained for the smooth model are
$\epsilon_- = \epsilon_+ = 0.00134884$, $\ell_0=2.416$, $K_0=0.008293$,
and $r_c=7.511$. We assume for consistency that the source is located at
the same radius $r =19.917$ in both the smooth and shocked cases.

The results for the specific angular momentum $\ell$ are significantly
sub-Keplerian, as expected based on the relatively high value utilized
for the viscosity parameter, $\alpha = 0.1$ (cf. Fig.~8{\it b}). The
sound speed is lower when a shock is present, due to the escape of
energy at the shock location, as noted previously in Figure~8{\it a}.
The result for $\Qjet$ obtained by substituting our values for
$\epsilon_+$ and $\epsilon_-$ into equation~(\ref{eq35}) is equal to the
observed jet luminosity, $5 \times 10^{38}\,{\rm erg \ s^{-1}}$, and
this establishes that our shocked disk model successfully accounts for
the observed energetics of the accretion flow and the outflow in \SgrA.

\section{CONCLUSION}

In this paper we have presented the first systematic study of the
structure of stationary, advection-dominated, shocked viscous accretion
disks around black holes. Our primary focus here has been on the global
properties of accretion flows containing isothermal shock waves, formed
as a result of the presence of a centrifugal barrier located near the
event horizon. Our model is based on the same set of standard ADAF
conservation equations considered by Narayan et al. (1997). The
conservation equations are supplemented by the inner boundary conditions
developed by Becker \& Le (2003) which allow the calculation of starting
values for the physical variables close to the horizon. Once the
starting values are established, a stable outward integration method is
employed to obtain the radial profiles of the various flow quantities
such as velocity, sound speed, and angular velocity. The fundamental
model parameters are the energy inflow rate at the horizon,
$\epsilon_0$, the specific angular momentum of the material at the
horizon, $\ell_0$, the entropy parameter at the horizon, $K_0$, the
Shakura-Sunyaev viscosity parameter, $\alpha$, and the ratio of specific
heats, $\gamma$.

For given observational values of the black hole mass $M$, the accretion
rate $\dot M$, and the jet kinetic luminosity $\Qjet$, we are free to
select values for $\alpha$, $\gamma$, and $\epsilon_0$. Following
Narayan et al. (1997), we have set $\gamma=1.5$ in the calculations
presented here. The remaining parameters $\ell_0$ and $K_0$, and the
shock radius $r_*$, are subsequently determined by applying the
asymptotic inner boundary conditions given by equations~(\ref{eq30}) and
(\ref{eq31}), the critical conditions given by equations~(\ref{eq24})
and (\ref{eq25}), and the shock jump conditions given by
equations~(\ref{eq34}) and (\ref{eq35}). The dynamical structure of the
disk/shock system is obtained by numerically integrating
equations~(\ref{eq11}), (\ref{eq15}), and (\ref{eq20}) in the outward
direction beginning at a starting radius located close to the event
horizon.

The formation of a shock requires that the flow possess at least two
critical points. The existence of multiple critical points in both
adiabatic and polytropic accretion flows has been reported by many
authors (e.g., Abramowicz \& Zurek 1981; Chakrabarti 1989a, 1989b, 1990,
1996). Flows with multiple critical points are confined to a restricted
region of the parameter space, as indicated in Figure~7. Accretion flows
within this region can form isothermal shocks. In their earlier study,
Narayan et al. (1997) focused solely on solutions that pass only through
the inner critical point, and consequently they did not treat the
shocked disk solutions. Flows with discontinuities must first pass
through an outer critical point, and subsequently through an isothermal
shock and an inner critical point in order to represent physically
acceptable accretion solutions. That is the scenario we have
investigated here.

\begin{figure}
\centering
\includegraphics[scale=0.7]{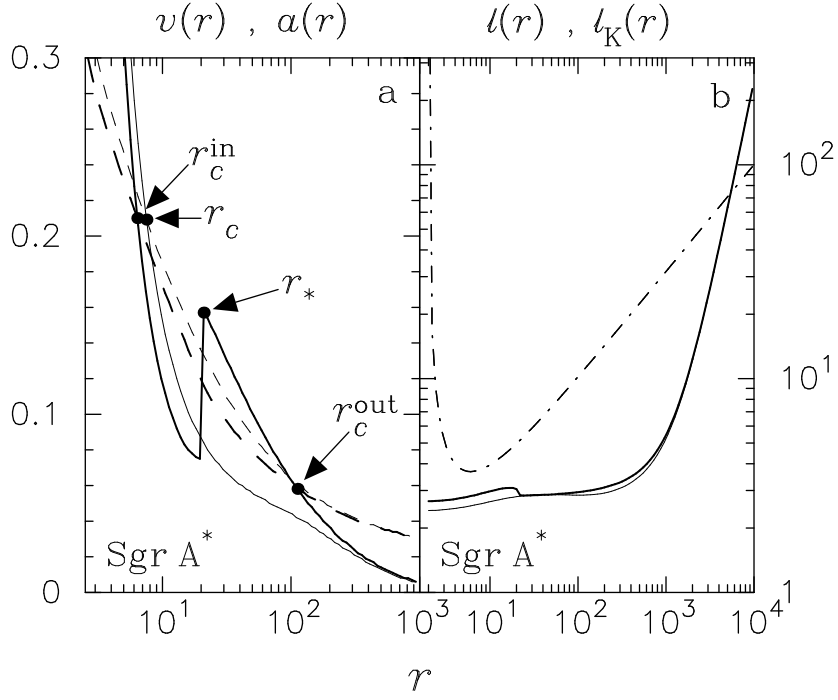}
\caption{Same as Fig.~8, except model results are for \SgrA.}
\end{figure}

The possible presence of shock waves in accretion disks and their
observational implications have been examined by a variety of previous
authors. Abramowicz \& Chakrabarti (1990) presented one of the earliest
studies of the properties of standing shock waves in accretion disks.
Subsequently, the particular importance of isothermal shock waves was
realized by Le \& Becker (2004, 2005, 2007), who pointed out that
efficient particle acceleration in the vicinity of a standing isothermal
shock can help to explain the formation of the relativistic outflows of
matter commonly observed around radio-loud, underfed black holes.
However, all of these previous studies of isothermal shocks in accretion
disks neglected the important role of viscosity in determining the disk
structure. Very recently, Becker, Das, \& Le (2008) demonstrated that
first-order Fermi acceleration inside a viscous, shocked ADAF can
accelerate particles more efficiently than in a smooth disk. Hence the
presence of a standing shock wave can create a potentially favorable
environment for the production of relativistic outflows. In the present
paper, we have presented a detailed study of the dynamics of viscous,
shocked ADAF disks that are governed by the standard description of the
angular momentum transport based on the Shakura-Sunyaev prescription for
the kinematic viscosity. In particular, we confirm that the energetics
of the viscous disk/shock model can explain the observational properties
of both the accretion flow and the relativistic outflows in M87 and \SgrA.

In Figure~7 we explored the parameter space within which ADAF disks with
isothermal shocks can exist as a function of the specific angular
momentum at the horizon $\ell_0$ and the incident energy transport rate
$\epsilon_-$ for various (constant) values of the viscosity parameter
$\alpha$. The shock region shifts toward lower values of $\ell_0$ as
$\alpha$ increases due to the enhanced outward diffusion of angular
momentum. We find that for the canonical value $\gamma=1.5$, shocked
disk solutions can exist provided $\alpha \lapprox 0.27$. This
establishes for the first time that highly viscous ADAF disks can
contain standing shock waves. The stability of such a shock is an open
question, but Nagakura \& Yamada (2008) and Okuda et al. (2007) recently
found that shocks in inviscid disks are persistent, although the shock
radius may oscillate. These results at least suggest the possibility
that shocks in viscous disk may be stable as well, although this needs
to be investigated in future work. Since shocked solutions are expected
to possess higher entropy than smooth solutions, we argue based on the
second law of thermodynamics that when shock solutions are permitted
dynamically, they should be expected to form (Becker \& Kazanas 2001).

Although Gu \& Lu (2001, 2004) obtained solutions for ADAF disks
containing Rankine-Hugoniot shocks, the results developed here represent
the first dynamical solutions for ADAF disks with a significant level of
viscosity (i.e., $\alpha=0.1$) containing isothermal shocks. The
Rankine-Hugoniot shocks studied by Gu \& Lu (2001, 2004) have conserved
energy transport rates, and therefore they cannot directly produce
relativistic outflows (jets). Conversely, the isothermal shocks we study
here have a discontinuous energy transport rate and consequently they
are able to power the observed outflows. In our analysis of the disk
structure, we have assumed that the accretion rate $\dot M$ remains
constant throughout the disk, and therefore we have neglected the effect
of the outflow mass loss on the dynamics of the disk, despite the fact
that an outflow emanates from the shock location in our model. This
assumption is probably reasonable since estimates show that the rate of
mass loss from the shock into the wind is typically $\sim 10^{-3}\,\dot
M$ if the jet is hadronic (Le \& Becker 2005); this ratio is further
reduced by the factor $m_e/m_p$ if the jet is leptonic. However, there
are two other potential forms of mass loss from the disk that must also
be considered. The first is the possibility of hydrodynamically driven
outflows of the background thermal gas, and the second is the
possibility of the escape of a significant number of relativistic
particles from the inner and/or outer regions of the disk, rather than
from the shock itself. We consider each of these issues below.

Narayan et al. (1997) and Blandford \& Begelman (1999) have noted
that ADAF disks typically possess a positive Bernoulli parameter,
\begeq
B(r) \equiv {v^2 \over 2} + {\ell^2 \over 2 \, r^2} + {\gamma
\, a^2 \over \gamma-1} - {GM \over r-\rs} \ ,
\label{eq36}
\fineq
suggesting that sufficient energy is available to power hydrodynamical
outflows of the thermal gas. It is interesting to reexamine this issue
in the context of the shocked, viscous accretion disks considered here.
In Figures~10{\it a} and 10{\it b}, respectively, we plot the Bernoulli
parameter as a function of radius based on the smooth and shocked models
for M87 and \SgrA~developed in \S~5. Both the smooth and shocked
solutions display a slightly positive Bernoulli parameter in the inner
and outer regions, suggesting that the gas is unbound there. When a
shock is present, the gas is bound in the post-shock region, but it
becomes unbound again before crossing the event horizon.

The Bernoulli parameter is related to the energy transport rate $\dot E$
and the viscous torque ${\cal G}$ via (see eq.~[\ref{eq3}])
\begeq
\dot E = - \mathcal{G} \, \Omega +
\dot M B \ .
\label{eq37}
\fineq
Since $\dot E$ is conserved in the inner region, it follows that the
positive bump in the Bernoulli parameter between the shock and the
horizon reflects an increase in the torque-driven outward energy
transport rate, $\mathcal{G} \, \Omega$. This effect is less pronounced
in the shocked solution because the escaping particles carry away a
significant fraction of the binding energy. The positivity of the
Bernoulli parameter in the inner and outer regions suggests the
possibility of hydrodynamical outflows emanating from the non-shock
regions of the flow. However, any such outflows would be quite weak,
with terminal Lorentz factors $\Gamma = B + 1 \sim 1.01$, whereas the
terminal Lorentz factor for the outflow from the shock is $\Gamma \sim
6$ (Becker et al. 2008; 2009). More fundamentally, it must be emphasized
that despite the positivity of the Bernoulli parameter, there is no
hydrodynamical mechanism in the model considered here capable of
generating an outflow, since the ADAF disks we focus on are in vertical
hydrostatic equilibrium, and therefore the net vertical force vanishes
(cf. eq.~[\ref{eq13}]). Hence some additional mechanism, outside the
standard ADAF framework, must be operative in order to channel the
energy into the outflows. In our view, the outflows are not
hydrodynamical in nature, but are instead powered by the first-order
Fermi acceleration of relativistic particles in the vicinity of the
discontinuous shock.

In addition to the possibility of hydrodynamical mass loss, some mass
may also be lost from the disk in the form of relativistic particles
that escape from the inner and out regions of the disk, rather than from
the shock itself. The escape of these additional relativistic particles
could be driven by the distributed first-order Fermi acceleration
occurring throughout the disk as a consequence of the overall
compression of the flow. The significance of this effect can only be
understood by performing a detailed study of the relativistic particle
transport occurring throughout the disk, which is beyond the scope of
the present paper. We are currently developing such a model and will
report the results in a separate paper. Preliminary calculations suggest
that the spatial distribution of the escaping particles is strongly
peaked at the shock location, with about half of the particles escaping
from the shock itself, and the remainder escaping from the surrounding
region (Becker et al. 2009). Hence, we conclude that the loss of mass
from the disk is strongly concentrated in the vicinity of the shock, and
that it is dominated by the escape of relativistic particles, rather
than by hydrodynamical outflows of the background (thermal) gas.

\begin{figure}
\centering
\includegraphics[scale=0.7]{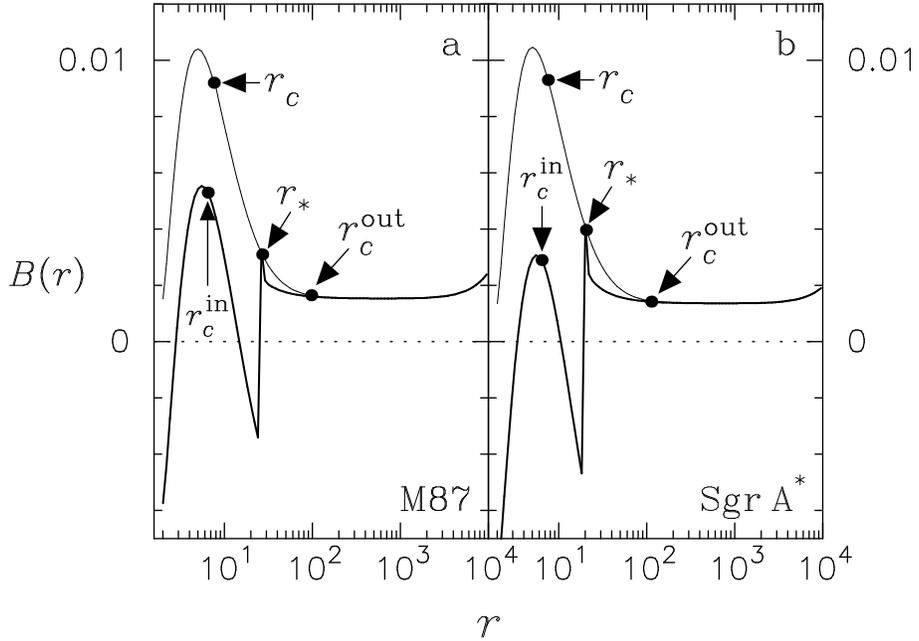} \caption{Bernoulli parameter
(eq.~[\ref{eq36}]) plotted as a function of the radius $r$ in units of
$GM/c^2$ for ({\it a}) M87 and ({\it b}) \SgrA. The thick and thin lines
denote the shocked and smooth solutions, respectively.}
\end{figure}

The authors would like to thank the anonymous referee for providing
several useful comments that significantly strengthened the paper. SD
acknowledges support via a postdoctoral fellowship from the Korea
Astronomy and Space Science Institute (KASI).

{}

\label{lastpage}


\clearpage

\begin{thebibliography}{}

\bibitem{1} Abramowicz, M. A., \& Chakrabarti, S. K. 1990, \apj, 350, 281

\bibitem{2} Abramowicz, M. A., \& Zurek, W. H. 1981, \apj, 246, 314

\bibitem{3} Becker, P. A., Das, S., \& Le, T. 2008, \apj, 677, L93

\bibitem{4} Becker, P. A., Das, S., \& Le, T. 2009, in preparation

\bibitem{5} Becker, P. A., \& Kazanas, D. 2001, \apj, 546, 429

\bibitem{6} Becker, P. A., \& Le, T. 2003, \apj, 588, 408

\bibitem{7} Becker, P. A., \& Subramanian, P. 2005, \apj, 622, 520

\bibitem{8} Bicknell, G. V., \& Begelman, M. C. 1996, \apj, 467, 597

\bibitem{9} Blandford, R. D., \& Begelman, M. C. 1999, \mnras, 303, L1

\bibitem{10} Caditz, D. M., \& Tsuruta, S. 1998, \apj, 501, 242

\bibitem{11} Chakrabarti, S. K. 1989a, Publ. Astron. Soc. Japan, 41, 1145

\bibitem{12} Chakrabarti, S. K. 1989b, \apj, 347, 365

\bibitem{13} Chakrabarti, S. K. 1990, Theory of Transonic Astrophysical Flows. 
World Scientific Publishing, Singapore

\bibitem{14} Chakrabarti, S. K. 1996, \apj, 464, 664

\bibitem{15} Chakrabarti, S. K., Acharyya, K., \& Molteni, D. 2004,
A\&A, 421, 1

\bibitem{16} Chakrabarti, S. K., \& Das, S. 2004, \mnras, 349, 649

\bibitem{17} Chakrabarti, S. K., \& Molteni, D. 1993, \apj, 417, 671

\bibitem{18} Das, S. 2007, \mnras, 376, 1659

\bibitem{19} Das, S., \& Chakrabarti, S. K. 2004, IJMPD, 13, 1955

\bibitem{20} Das, S., Chattopadhyay, I., \& Chakrabarti, S. K. 2001,
\apj, 557, 983

\bibitem{21} Das, T. K., Pendharkar, J. K., \& Mitra S. 2003, \apj, 592, 1078

\bibitem{22} Falcke, H., \& Biermann, P. L. 1999, A\&A, 342, 49

\bibitem{23} Ford, H. C., et al. 1994, \apj, 435, L27

\bibitem{24} Frank, J., King, A. R., \& Raine, D.~J. 2002, Accretion
Power in Astrophysics (Cambridge: Cambridge University Press)

\bibitem{25} Fukue, J. 1987, PASJ, 39, 309

\bibitem{26} Fukumura, K., \& Tsuruta, S. 2004, \apj, 611, 964

\bibitem{27} Gu, W.-M., \& Lu, J.-F. 2001, Chin. Phys. Lett., 18, 148

\bibitem{28} Gu, W.-M., \& Lu, J.-F. 2004, Chin. Phys. Lett., 21, 2551

\bibitem{29} Hawley, J. F., Smarr, L. L., \& Wilson, J. R. 1984a, \apj,
277, 296

\bibitem{30} Hawley, J. F., Smarr, L. L., \& Wilson, J. R. 1984b,
\apjs, 55, 211

\bibitem{31} Kovalenko, I. G., \& Lukin, D. V. 1999, Astron. Let., 25, 215

\bibitem{32} Le, T., \& Becker, P. A. 2004, \apj L, 617, L25

\bibitem{33} Le, T., \& Becker, P. A. 2005, \apj, 632, 476

\bibitem{34} Le, T., \& Becker, P. A. 2007, \apj, 661, 416

\bibitem{35} Lightman A. P., \& Eardley D. M. 1974, \apj, 187, L1

\bibitem{36} Lu, J. F., \& Yuan, F. 1998, \mnras, 295, 66

\bibitem{37} Ma, R.-Y., Yuan, F., \& Wang, D.-X. 2007, \apj, 671, 1981

\bibitem{38} Molteni, D., Sponholz, H., \& Chakrabarti, S. K. 1996,
\apj, 457, 805

\bibitem{39} Muchotrzeb, B., \& Paczy\'nski B. 1982, Acta Astron., 32, 1

\bibitem{40} Nagakura, H., \& Yamada, S. 2008, \apj, 689, 391

\bibitem{41} Narayan, R., Kato, S., \& Honma, F. 1997, \apj, 476, 49

\bibitem{42} Narayan, R., \& McClintock, J. E. 2008, NewAR, 51, 733

\bibitem{43} Nobuta K., \& Hanawa T. 1994, Publ. Astron. Soc. Japan, 46, 257

\bibitem{44} Novikov I., \& Thorne K. S. 1973, Black Holes, Eds. C. DeWitt
and B. DeWitt (Gordon and Breach, New York)

\bibitem{45} Okuda, T., Teresi, V., \& Molteni, D. 2007, \mnras, 377, 1431

\bibitem{46} Owen, F. N., Eilek, J. A., \& Kassim, N. E. 2000, \apj,
543, 611

\bibitem{47} Paczy\'nski B., \& Bisnovatyi-Kogan G. 1981, Acta Astron., 31, 283

\bibitem{48} Paczy\'nski, B., \& Wiita, P. J. 1980, A\&A, 88, 23

\bibitem{49} Quataert, E. 2003, Astron. Nachr., 324, 435

\bibitem{50} Reif, F. 1965, Fundamentals of Statistical and Thermal
Physics (NY: McGraw-Hill)

\bibitem{51} Reynolds, C. S., et al. 1996, MNRAS, 283, L111

\bibitem{52} Ryu, D., Chakrabarti, S. K., \& Molteni, D. 1997, \apj, 474, 378

\bibitem{53} Sch\"odel, R., et al. 2002, Nature, 419, 694

\bibitem{54} Shakura, N. I., \& Sunyaev. R. A. 1973, A\&A, 24, 337

\bibitem{55} Takahashi, R. 2007, A\&A, 461, 393

\bibitem{56} Weinberg, S. 1972, Gravitation and Cosmology (NY: Wiley)

\bibitem{57} Yang, R., \& Kafatos, M. 1995, A\&A, 295, 238

\bibitem{58} Yu, C., Lou, Y. Q., Bian, F. Y., \& Wu, Y. 2006, \mnras, 370, 121

\bibitem{59} Yuan, F. 2000, MNRAS, 319, 1178

\bibitem{60} Yuan, F. 2007, ASPC, 373, 95

\bibitem{61} Yuan, F., Dong, S., \& Lu, J. F. 1996, Ap\&SS, 246, 197

\bibitem{62} Yuan, F., Ma, R., \& Narayan, R. 2008, \apj, 679, 984

\bibitem{63} Yuan, F., Markoff, S., \& Falcke, H. 2002, A\&A, 383, 854

\end{thebibliography}
\end{document}